\documentclass[twocolumn,longbibliography,superscriptaddress,prd,showpacs]{revtex4-1}

\usepackage{amsmath}
\usepackage{amssymb}
\usepackage{xcolor}

\usepackage{bm}

\usepackage{amsmath}
\usepackage{amssymb}

\newcommand{\dd}{{\mathrm d}}

\newcommand{\ee}{{\mathrm e}}

\newcommand{\calL}{{\mathcal L}}

\begin{document}

\title{Atomic Physics Constraints on the $\bm{X}$ Boson}

\author{Ulrich D. Jentschura}

\affiliation{Department of Physics,
Missouri University of Science and Technology,
Rolla, Missouri 65409-0640, USA}

\affiliation{MTA--DE Particle Physics Research Group,
P.O.Box 51, H--4001 Debrecen, Hungary}

\affiliation{MTA Atomki, P.O.Box 51, H--4001 Debrecen, Hungary} 

\author{Istv\'{a}n N\'{a}ndori}

\affiliation{MTA--DE Particle Physics Research Group,
P.O.Box 51, H--4001 Debrecen, Hungary}

\affiliation{MTA Atomki, P.O.Box 51, H--4001 Debrecen, Hungary} 

\affiliation{Department of Theoretical Physics, 
University of Debrecen, P.O.Box 105, Debrecen, Hungary}

\begin{abstract}
Recently, a peak in the light fermion pair spectrum 
at invariant $q^2 \approx (16.7 \, {\rm MeV})^2$
has been observed in the bombardment of ${}^7 {\rm Li}$
by protons. This peak has been interpreted in terms of a 
protophobic interaction of fermions with a 
gauge boson ($X$ boson) of invariant mass $\approx 16.7 \, {\rm MeV}$
which couples mainly to neutrons.
High-precision atomic physics experiments 
aimed at observing the protophobic interaction
need to separate the $X$ boson effect 
from the nuclear-size effect, which is a problem
because of the short range
of the interaction ($11.8 \, {\rm fm}$),
which is commensurate with a  ``nuclear halo''.
Here, we analyze the $X$ boson in terms of its consequences
for both electronic atoms as well as 
muonic hydrogen and deuterium.
We find that the most promising atomic systems where 
the $X$ boson has an appreciable effect, distinguishable
from a finite-nuclear-size effect, are
muonic atoms of low and intermediate nuclear charge numbers.
\end{abstract}


\maketitle

%
%
\section{Introduction}
\label{sec1}

Recently, the reaction
\begin{equation} 
{}^7 {\rm Li} + p \to
{}^8 {\rm Be}^* \to 
{}^8 {\rm Be} + \gamma \to 
{}^8 {\rm Be} + e^+ e^-
\end{equation}
has been observed at the MTA ATOMKI (Institute 
for Nuclear Research of the Hungarian Academy of Sciences) in Debrecen,
and deviations from Standard Model predictions have 
been recorded~\cite{KrEtAl2016,*KrEtAl2017woc2,Ca2016news,Ta2016}.
While the primary aim of the 
study had been the hunt for a massive (``dark'')
photon, the experimental data are described satisfactorily 
in terms of a new (``fifth-force'') $X$ boson 
(vector boson) which couples
to fermions according to~\cite{FeEtAl2016,FeEtAl2017}
\begin{equation}
\label{LL}
\calL = - e \, \sum_f \varepsilon_f \, 
\overline \psi_f \, \gamma_\mu \, X^\mu \, \psi_f \,,
\end{equation}
where $X^\mu$ is the spin-1 $X$ boson field, 
$f$ sums over the fermions (fermion flavors),
and the $\varepsilon_f$ coefficients 
describe the flavor-dependent
couplings to the $X$ boson.
A family-dependence is disfavored by the authors
of Refs.~\cite{FeEtAl2016,FeEtAl2017}.
Rather, the $X$ boson is advocated as a possible partial
explanation for the observed $3.6 \sigma$ discrepancy of the
observed muon $g$ factor~\cite{Po2009},
while assuming a family independence 
(electron versus muon) of the couplings $\varepsilon_f$
(i.e., in particular, $\varepsilon_e \approx \varepsilon_\mu$
for electron and muon).

If, accidentally, the following combination of couplings
to the up and down quarks add to a value close zero,
\begin{equation} 
\label{protophobic}
2 \, \varepsilon_u + \varepsilon_d \approx 0 \,,
\end{equation}
then the interaction with the $X$ boson becomes protophobic,
i.e., protons are effectively decoupled.
By contrast, a numerical value of 
\begin{equation} 
|\varepsilon_n| = |\varepsilon_u + 2 \varepsilon_d |
\approx \left| \frac32 \, \varepsilon_d \right| 
\approx \frac{1}{100}
\end{equation}
explains the observed $6.1\sigma$ peak seen in the 
experiments~\cite{KrEtAl2016,*KrEtAl2017woc2,Ca2016news,Ta2016,FeEtAl2016,FeEtAl2017}
[see Eq.~(10) of Ref.~\cite{FeEtAl2016}].
The proposed vector boson has a mass of $m_X = 16.7 \, {\rm MeV/c^2}$.
Light particles similar to dark photons in this mass 
range have been considered a possible solution 
to problems related to the understanding of certain 
isotope abundances in the Universe~\cite{GoPoPr2016},
and other experiments have been designed to cover 
the conjectured parameter range of the $X$ boson~\cite{EcEsZh2015}
(for a more detailed discussion of the 
particle physics aspects of the proposed boson, see the Appendix).

From below, the parameter $\varepsilon_e$ for the 
electrons is further constrained by electron beam dump 
experiments, which search for dark photons~\cite{FeEtAl2016,FeEtAl2017},
while a high bound on $\varepsilon_e$ is set
by electron $g-2$ experiments. Numerically, one finds 
that~\cite{FeEtAl2016,FeEtAl2017}, 
\begin{equation}
\label{epsEbound}
2 \times 10^{-4} < \varepsilon_e < 1.4 \times 10^{-3} \,.
\end{equation}
Traditionally, atomic high-precision experiments have 
been used with good effect to constrain any conjectured 
additions to the low-energy sector of the 
Standard Model (see, e.g., Refs.~\cite{JaRo2010,Je2011aop2}). 
Moreover, it has been one of the goals of
high-precision atomic spectroscopy to explore the 
low-energy sector of the Standard Model,
and to possibly discover a ``hidden'' sector 
of fundamental interactions at low energy~\cite{Ha2006}.
Several recent papers 
explore the consequences of the proposed $X$ boson 
for atomic spectroscopy, notably, isotope 
shifts~\cite{BeEtAl2017darkphoton,FrFuPeSc2017,MiTaYa2017}.
The purpose of the current paper is twofold.
First, we briefly discuss 
possible implications of the $X$ boson
for the proton and deuteron radius puzzle, which still
has not been completely solved~\cite{BeEtAl2017darkphoton,Fl2017,*FlEtAl2017}
(see Sec.~\ref{sec3}).
Second, we attempt to find a simple atomic system, 
in which the effect of the $X$ boson could be discerned, 
based on a straightforward theoretical analysis,
without resorting to numerical many-body calculations
of isotope shifts~\cite{BeEtAl2017darkphoton,FrFuPeSc2017,MiTaYa2017}
(see the discussion in Sec.~\ref{sec5}).

Also, we shall attempt to develop an intuitive understanding
for the observation~\cite{MiTaYa2017} that it is rather difficult
to obtain a signal from the $X$ boson in electronic 
bound systems (as discussed in Sec.~\ref{sec2}).
A promising alternative appears to involve
muonic systems with medium and high nuclear charge
numbers, for reasons to be discussed in the 
following. 

%
%
\section{Energy Scales}
\label{sec2}

In order to obtain a somewhat intuitive understanding
of the $X$ boson in terms of atomic physics,
it is instructive to explore the energy scales 
involved in the problem. Indeed,
the proposed vector boson mass of $m_X = 16.7 \, {\rm MeV}/c^2$
is much larger than both the effective mass $\alpha \, m_e$
of bound electronic systems,
as well as the momentum scale 
$\langle p \rangle = Z \, \alpha \, m_e \, c \approx  0.343 \, {\rm MeV}/c$
of hydrogenlike Uranium ($Z = 92$),
and also larger than the bound-state momentum
$\langle p \rangle = \alpha \, m_\mu \, c\approx 0.772 \, {\rm MeV}/c$ 
of muonic hydrogen~\cite{PoEtAl2010},
but not necessarily larger than the 
momentum scale $\langle p \rangle = Z \, \alpha m_\mu c$ of a 
one-muon ion with medium charge number $Z$.
E.g., for muonic carbon, one has a momentum scale
$\langle p \rangle = 6 \, \alpha m_\mu \, c \approx 4.63 \, {\rm MeV}/c$ 
which is commensurate with the $X$ boson mass.
For muonic magnesium, one has 
$\langle p \rangle = 12 \, \alpha m_\mu \, c \approx 9.25 \, {\rm MeV}/c$.
These considerations are relevant because the $X$ boson mass
determines the range of the interaction 
mediated by the new particle, which is
$\langle r \rangle = 
\hbar/\langle p \rangle$.

For electronic systems, the energy scale of the 
$X$ boson is ``detached'' from both 
electronic bound systems as well as low-$Z$
muonic bound systems.
The range of the $X$ boson interaction is equal to 
its reduced Compton wavelength,
\begin{equation}
\label{LX}
\lambdabar_X = \frac{\hbar}{m_X c} = 11.8 \, {\rm fm} \,,
\end{equation}
which has to be compared to the generalized Bohr radius for
muonic hydrogen,
\begin{equation}
\label{LmuH}
\lambdabar_{\mu{\rm H}} = 
\frac{\hbar}{\alpha m_\mu c} = 256\, {\rm fm} \,,
\end{equation}
and the (ordinary) hydrogen atom,
\begin{equation}
\label{LH}
\lambdabar_H = 
\frac{\hbar}{\alpha m_e c} = a_0 = 52917.7 \, {\rm fm}  \,,
\end{equation}
where $a_0$ is the (ordinary) Bohr radius.
As already indicated, 
the Bohr radius for a one-muon carbon ion,
\begin{equation}
\label{L12C}
\lambdabar_{\mu {}^{12}{\rm C}} = 
\frac{\hbar}{6 \, \alpha \, m_\mu \, c} 
= 42.6 \, {\rm fm} \,,
\end{equation}
is closer to the range of the $X$ boson interaction.
In the following, we refer to the bound system with a 
single, negatively charged muon circling around a carbon nucleus,
as ``muonic carbon''. For muonic magnesium, as defined analogously,
we have $\lambdabar_{\mu {}^{24}{\rm Mg}} = 21.3 \, {\rm fm}$
(with nuclear charge number $Z=12$).

From now on, we shall use natural units with 
$\hbar = c = \varepsilon_0 = 1$.
By matching the scattering amplitude 
generated by the Lagrangian~\eqref{LL}
to an effective Hamiltonian in the no-retardation 
approximation (zero energy of the virtual boson), 
we obtain the following 
interaction Hamiltonian 
$H_X$ for electronic bound systems 
(in the low-energy limit),
\begin{subequations}
\label{HX}
\begin{equation}
\label{HXe}
H^{(e)}_X = \varepsilon_e \, 
\varepsilon_n \, (A-Z) \, (4 \pi \alpha) \, 
\frac{\delta^{(3)}(\vec r)}{m_X^2} \,.
\end{equation}
Here, $A$ is the mass number of the nucleus, while
$Z$ is the charge number, so that
$A-Z$ counts the number of neutrons in the 
nucleus. If the orbiting particle 
is a muon, then we need to replace
$\varepsilon_e \to \varepsilon_\mu$ and obtain
\begin{equation}
\label{HXmu}
H^{(\mu)}_X = \varepsilon_\mu \, 
\varepsilon_n \, (A-Z) \, (4 \pi \alpha) \, 
\frac{\delta^{(3)}(\vec r)}{m_X^2} \,.
\end{equation}
\end{subequations}
The finite-nuclear-size (FNS) Hamiltonian is~\cite{Je2003jpa}
\begin{equation}
\label{HFNS}
H_{\rm FNS} = \frac{2 \pi}{3} 
Z \alpha \; r^2_n \;
\delta^{(3)}(\vec r) \,,
\end{equation}
where $r_n = \sqrt{ \left< r^2_n \right> }$ 
is the root-mean-square 
charge radius of the nucleus.
The two Hamiltonians~\eqref{HX} and~\eqref{HFNS}
are both proportional to a Dirac-$\delta$ function. 

%
%
\section{$\bm X$ Boson and Deuteron Radius}
\label{sec3}

Let us explore a possible role of the $X$ boson
in the proton and deuteron radius puzzle~\cite{PoEtAl2010,AnEtAl2013,PoEtAl2016},
and take into account a possible 
family dependence of the interaction, 
i.e., ask the question of whether a coupling constant
dependence $\varepsilon_e \neq \varepsilon_\mu$
could contribute to an explanation of the puzzle.
The current status of this puzzle can be 
summarized as follows:
For the proton, a recent measurement~\cite{BeEtAl2017}
of the $2S$--$4P$ transition
has indicated a possible reconciliation,
by analyzing a cross-damping term 
(``nonresonant shift'') of the 
transition due to neighboring fine-structure 
states~\cite{JeMo2002}.
The revised value of the proton radius~\cite{BeEtAl2017}, 
derived from hydrogen spectroscopy, is 
$r_p = 0.8335(95)\, {\rm fm}$
and in better agreement
with the muonic hydrogen value $r_p = 
0.84087(39) \, {\rm fm}$
than the previous CODATA value of 
$r_p = 0.8775(51) \, {\rm fm}$, which is
primarily derived from an analysis of the most accurately 
measured hydrogen transitions 
(see Table~XXXVIII of Ref.~\cite{MoTaNe2012}).
One notes that the ``larger'' proton radius of 
$r_p \approx 0.88 \, {\rm fm}$
is mainly derived in combining very accurate
$1S$--$2S$ measurements~\cite{MaEtAl2013prl}
with $2S$--$nD$ measurements~\cite{BeEtAl1997,ScEtAl1999}
and $1S$--$3S$ atomic hydrogen measurements~\cite{Fl2017}
of the Paris group. One might speculate about 
an incomplete analysis of the systematic effects
in the measurements of the Paris group; however,
a very recent work~\cite{Fl2017} reaffirms the correctness 
of the analysis performed for the $2S$--$8D$ and $2S$--$12D$
transitions, and $1S$--$3S$ transitions~\cite{Fl2017}. 
One can thus, at present, not conclusively confirm
that the proton radius puzzle has been solved.
In any case, for the proton, it turns out that 
the $X$ boson cannot contribute to an 
explanation of the puzzle, 
because of the protophobic character of the 
proposed interaction [see Eq.~\eqref{protophobic}].

For the deuteron, the CODATA value of
$r_d = 2.1424(21) \, {\rm fm}$ is
primarily derived from (ordinary) 
deuterium spectroscopy~\cite{PoEtAl2016}.
It has to be compared to the value
$r_d = 2.12562(78) \, {\rm fm}$ 
derived from muonic deuterium spectroscopy~\cite{PoEtAl2016}.
The relative difference of these values is 
\begin{equation}
\frac{\delta r_d^2}{r_d^2} = 0.016(2) \,.
\end{equation}
Let us assume, for the moment, that this 
difference is due to a lepton family
non-universality of the $X$ boson interaction. 
To this end, we evaluate the ratio of the 
energy shift due to the $X$ boson,
to the finite-size energy shift.
This ratio is equal to the ratio of the 
change $\delta r_d^2$ 
in the root-mean-square radii to the root-mean-square
charge radius of the deuteron itself,
\begin{equation}
\label{013}
\frac{\langle H^{(e)}_X - H^{(\mu)}_X \rangle}{\left< H_{\rm FNS} \right>} 
= \frac{6 \, (\varepsilon_e - \varepsilon_\mu) \, \varepsilon_n }%
{ m_X^2 \, r_n^2 } \, \frac{A-Z}{Z} 
= \frac{\delta r_d^2}{r_d^2} \,,
\end{equation}
where $A=2$, $Z=1$, $\varepsilon_n \approx 1/100$
[see Eq.~(32) of Ref.~\cite{FeEtAl2017}]. 
Plugging in values, one obtains
\begin{equation}
\label{012}
(\varepsilon_e - \varepsilon_\mu) \approx 0.012 \,.
\end{equation}
The sign can be understood from the fact that 
the conceivable existence of the $X$ boson, for electronic
systems, would enhance the finite-size Dirac-$\delta$ potential,
for $\varepsilon_e > 0$, and thus lead to a larger 
value of the deuteron radius, if determined from electronic 
bound systems.
The result~\eqref{012} is incompatible with the bound~\eqref{epsEbound}
for the coupling parameter of the electron, 
assuming an approximate family independence 
$\varepsilon_e \approx \varepsilon_\mu$ of the couplings.
Furthermore, assuming $\varepsilon_e \approx 0$, 
the value $\varepsilon_\mu = -0.012$ leads to a 
severe discrepancy with the muon $g-2$ experiment,
inducing a contribution to the muon anomaly
$(g-2)/2$ of about $1.58 \times 10^{-7}$ [see Eq.~(4) 
of Ref.~\cite{Po2009}]. The $X$ boson can thus be excluded 
as an explanation for the deuteron radius puzzle.

However, the conceivable existence of the $X$ boson would 
(slightly) affect the determination of the deuteron radius from experiments. 
Namely, one normally defines the
deuteron radius as the slope of the 
charge form factor $G_C$ of the deuteron
at zero momentum transfer,
after all QED effects and effects of 
``external'' interactions (virtual gauge bosons, etc.) 
have been subtracted [see Eq.~(13) of Ref.~\cite{AbEtAl2000}].
The slope of the charge form factor $G_C$ leads to the 
deuteron radius (see~\cite{GaVO2001} and Sec.~4.2 of 
Ref.~\cite{Ko2004deuteron})
\begin{equation}
r^2_d = 6 \, \left. \frac{\dd G_C(q^2)}{\dd q^2} \right|_{q^2 = 0} 
= -6 \, \left. \frac{\dd G_C(Q^2)}{\dd Q^2} \right|_{Q^2 = 0} \,,
\end{equation}
where $Q^2 = -q^2$ is the squared four-momentum transfer. 
Taking the $X$ boson into account,
the deuteron radius would shift
according to the replacements
\begin{equation}
\label{shift1}
r_d^2 \to r_d^2 - \frac{6 \, \varepsilon_\mu \, \varepsilon_n }{ m_X^2 } 
\end{equation}
for the determination from muonic deuterium,
and according to 
\begin{equation}
\label{shift2}
r_d^2 \to r_d^2 - \frac{6 \, \varepsilon_e \, \varepsilon_n }{ m_X^2 } 
\end{equation}
for determinations involving ordinary deuterium atoms.
Taking into account the bound~\eqref{epsEbound}
and assuming that $\varepsilon_e \approx \varepsilon_\mu$,
the shifts~\eqref{shift1} and~\eqref{shift2} are seen not to exceed 
$0.003 \, {\rm fm}$ when expressed in terms of the root-mean-square 
radius $r_d$.

Finally, let us note that the $X$ boson 
does not affect the 
determination of the Rydberg constant
from hydrogen and deuterium spectroscopy~\cite{JeKoLBMoTa2005}.
We recall that the Rydberg constant is one of the most 
accurately known physical constants, with a relative accuracy 
on the level of $10^{-12}$~\cite{MoTaNe2012,BeEtAl2017}.
However, one notes that the inclusion of the 
$X$ boson Hamiltonian~\eqref{HX} in the 
theoretical model for the determination of the 
Rydberg constant from hydrogen and deuterium spectroscopy 
would not affect the Rydberg constant,
because the additional term 
is of the same functional form as the 
finite-size Hamiltonian~\eqref{HFNS}
and thus reabsorbed in the nuclear radius.

%
%
\section{$\bm X$ Boson and Muonic Ions}
\label{sec4}

In principle, one might hope to determine the 
coupling parameter $\varepsilon_e$ from isotope shifts
of atomic transitions. 
The essential idea is to write the isotope shift 
as a linear combination of the mass shift of a
transition (due to the change in the reduced 
mass of the system), 
of the field shift (due to the isotopic change in the 
nuclear radius), and due to the $X$ boson
[see Eq.~(2) of Ref.~\cite{MiTaYa2017}].
We note that in principle, the mass shift 
could be obtained by very accurate Penning 
trap measurements and thus subtracted.
However, the observation of a single isotope shift 
does not determine the $X$ boson coupling because 
of the unknown field shift, i.e., the 
unknown radius difference. One might think that the 
radius could be determined independently 
by a scattering experiments and subtracted. 
However, in scattering experiments,
the $X$ boson term~\eqref{HX} modifies the 
scattering cross section just like the 
finite-nuclear-size term~\eqref{HFNS} and 
thus could not be subtracted separately.

Measurements of isotope shifts between the same isotopes
but more and different atomic transition also do not help because
in the leading-order approximation,
both the $X$ boson Hamiltonian~\eqref{HX} as well 
as the finite-size Hamiltonian ~\eqref{HFNS}  
are proportional to a Dirac-$\delta$.
One might observe isotope shifts 
involving more than two isotopes, 
considering that the prefactor of the $X$ boson term
depends on the isotope (via the change in the neutron number,
which enters the nuclear mass number $A$).
Even so, within the Dirac-$\delta$ approximation,
one still cannot accurately determine the $X$ boson coupling
because each addition 
of an isotope also implies the 
addition of a field shift term, i.e.,
an additional radius difference which cannot be determined 
independently. 

For electronic bound systems,
the reduced Compton wavelength $\lambdabar_H/(1 + n_e)$
[see Eq.~\eqref{LH}],
where $n_e$ is the charge number of the ion, 
is much larger than the reduced Compton wavelength
$\lambdabar_X$ of the $X$ boson, as given in
Eq.~\eqref{LX}. Thus, for electronic bound systems,
the $X$ boson potential remains a Dirac-$\delta$ 
to good approximation.
If at all, then the $X$ boson coupling
could be determined based on higher-order
terms beyond the Dirac--$\delta$ approximation 
used in Eqs.~\eqref{HX} and~\eqref{HFNS}
(see Refs.~\cite{FrFuPeSc2017,MiTaYa2017} for a comprehensive discussion,
especially in the context of ``King linearity violation''
as envisaged originally in Ref.~\cite{Ki1984}).
In the end, even under the optimistic 
assumption of an increase in the
precision of isotope spectroscopy to better than 
$1\, {\rm Hz}$, the range of coupling parameters 
and masses for the conjectured $X$ boson~\cite{FeEtAl2016,FeEtAl2017} 
remains out of the 
observable range of high-precision isotope shift 
measurements (specifically, see the black bar in the 
right panel Fig.~3.2 of Ref.~\cite{MiTaYa2017}).
A more optimistic point is taken by Ref.~\cite{FrFuPeSc2017},
where in Fig.~3, it is claimed that a measurement 
of isotope shifts in ${\rm Yb}^+$, 
involving nuclei with $A = 168, 170, 172, 174, 176$,
could potentially resolve the $X$ boson if an experimental 
accuracy of $1 \, {\rm Hz}$ is reached. This would 
correspond to an increase in the current level of 
experimental accuracy by four to five orders of magnitude.
Additionally, the drastic difference between the resolving power
of ${\rm Sr}^+$ and ${\rm Yb}^+$ reported 
in Fig.~3 of Ref.~\cite{FrFuPeSc2017} might be  considered as a little 
surprising because both ions have $n_e=1$, 
and so the reduced Compton wavelength (effective length 
scale of the atomic binding, effective nuclear charge number) 
is the same for the outer electrons in both systems. 
It would be somewhat awkward if the electron density 
in ${\rm Yb}^+$, which has a nuclear charge radius of 
about $5.3 \, {\rm fm}$~\cite{An2004} for the isotopes in question,
remains essentially constant over the nuclear volume,
while displaying a drastic deviation from the value 
inside the nucleus on a distance scale of $11.8 \, {\rm fm}$,
which is the range of the $X$ boson interaction.
Such a behavior would be required in order to substantially 
invalidate the Dirac--$\delta$ approximation used in Eqs.~\eqref{HX} and~\eqref{HFNS},
thus explaining the resolving power of isotope shifts 
in ${\rm Yb}^+$ as compared to ${\rm Sr}^+$, reported
in Fig.~3 of Ref.~\cite{FrFuPeSc2017}.
In any case, the precise understanding of the 
expansion coefficients 
used in Ref.~\cite{FrFuPeSc2017} may depend on the 
details of the many-body 
atomic structure code used in Ref.~\cite{FrFuPeSc2017}.

Here, we pursue a different route and attempt to find a
simple atomic system
where the $X$ boson contribution could naturally be extracted
based on a straightforward analytic model.
We need to find an atomic system
where the Dirac--$\delta$ approximation to the 
$X$ boson term~\eqref{HX} is insufficient,
and the $X$ boson Hamiltonian changes into
\begin{align}
\label{yukawaX}
H^{(e,Y)}_X =& \; \varepsilon_e \,  \varepsilon_n \,
(A-Z) \, \alpha \, \frac{\ee^{-m_X \, r}}{r} \,,
\\[0.1133ex]
\label{yukawaXb}
H^{(\mu,Y)}_X =& \; \varepsilon_\mu \,  \varepsilon_n \,
(A-Z) \, \alpha \, \frac{\ee^{-m_X \, r}}{r} \,.
\end{align}
Here, the superscript $Y$ reminds us of the
Yukawa character of the potential.
If the functional form of the $X$ boson term~\eqref{HX} 
and the finite-size term~\eqref{HFNS} are 
different for a particular atomic system, 
then we can distinguish the two effects.
For muonic carbon, according to Eq.~\eqref{L12C}, we have 
$\lambdabar_{\mu \, {}^{12}{\rm C}} = 42.6 \, {\rm fm}$,
which is commensurate with the reduced
Compton wavelength of the $X$ boson given in 
Eq.~\eqref{LX}, but much larger 
than the ${}^{12}{\rm C}$ radius of about $2.4 \, {\rm fm}$.
Hence, we have
\begin{equation}
\label{predestined}
r_{{}^{12} {\rm C}} \ll \lambdabar_{\mu \, {}^{12}{\rm C}} \,,
\qquad
\lambdabar_X \lesssim \lambdabar_{\mu \, {}^{12}{\rm C}} \,.
\end{equation}
This implies that in ${}^{12}{\rm C}$, 
the finite-nuclear-size 
Hamiltonian can still be approximated 
by a Dirac-$\delta$ potential,
while the $X$ boson Hamiltonian changes into 
the form given in Eq.~\eqref{yukawaXb}.

We note that at nuclear charge number $Z = 6$, 
one can still use nonrelativistic (Schr\"{o}dinger) 
wave functions to good approximation. 
In the relevant spectroscopic experiments on muonic 
carbon~\cite{ScEtAl1982,RuEtAl1984}
(for scattering data, see Ref.~\cite{OfEtAl1991}),
one observes the $1S$--$2P$ transition,
where the main nuclear-size effect is generated
by the expectation value of the finite-nuclear-size 
potential~\eqref{HFNS} in the ground state.
The ratio of the expectation values 
of the exact $X$ boson potential to the 
Dirac--$\delta$ approximation in the 
ground state is 
\begin{subequations}
\label{xi}
\begin{align}
\xi_{nS} =& \; \frac{ \langle nS | H^{(\mu, Y)}_X | nS \rangle }%
{ \langle nS | H^{(\mu)}_X | nS \rangle } \,,
\\[0.1133ex]
\xi_{1S} =& \; \frac{\chi^2}{(\chi + 2)^2} \,,
\\[0.1133ex]
\xi_{2S} =& \; \frac{\chi^2 \, (1 + 2 \, \chi^2)}{2 \, (1 + \chi)^2} \,,
\\[0.1133ex]
\label{xi3S}
\xi_{3S} =& \; \frac{3 \, \chi^2 \, \left[ 16 + 27 \, \chi^2 \, 
(8 + 9 \chi^2) \right]}{(2 + 3 \, \chi)^6} \,,
\end{align}
\end{subequations}
where $\chi$ is the ratio of the generalized Bohr radius
to the reduced Compton wavelength of the $X$ boson,
\begin{equation}
\label{defchi}
\chi = \frac{\lambdabar_{\mu \, {}^{12}{\rm C}}}{\lambdabar_X} =
\frac{m_X}{6 \alpha \, m_\mu} \approx 3.610 \,,
\end{equation}
and so we have $\xi_{1S} = 0.4140$, $\xi_{2S} = 0.3904$, and
$\xi_{3S} = 0.3865$. Of course, we have $\xi_{nS} \to 1$ for 
$\chi \to \infty$ ($m_X \to \infty$). In momentum space, 
the suppression of the correction for muonic 
carbon can be traced to the importance of 
spatial exchange momenta in excess of $m_X$,
which are important in the Coulomb exchange 
in the discussed atomic system.

In the 1970s, there has been some discussion regarding a
possible discrepancy in the determination of the 
charge radius of the ${}^{12}{\rm C}$ nucleus,
with values from electron scattering (without 
dispersion corrections) converging to a root-mean-square value
of $r_{{}^{12}{\rm C}} = 2.471(6) \, {\rm fm}$~\cite{Fr1979proc,OfEtAl1991},
while muonic spectroscopy led to a value of
$r_{{}^{12}{\rm C}} = 2.4829(19) \, {\rm fm}$~\cite{RuEtAl1984,OfEtAl1991}.
Under the assumption that this discrepancy is due to $X$ boson,
an analysis similar to the one carried out 
in Eqs.~\eqref{013} and~\eqref{012} leads to a value of 
\begin{equation}
(\varepsilon_\mu - \varepsilon_e) = 0.0070(46) \,,
\end{equation}
which deviates from zero by more than one standard deviation.
However, the large absolute magnitude of the required coupling 
coefficients excludes the $X$ boson as a viable 
explanation for the carbon charge radius discrepancy.
After the (somewhat ad hoc) application of dispersion corrections
to the scattering data, the 
value as determined from scattering has been shifted to
$r_{{}^{12}{\rm C}} = 2.478(9) \, {\rm fm}$~\cite{OfEtAl1991},
corresponding to 
\begin{equation}
(\varepsilon_\mu - \varepsilon_e) = 0.0029(64) \,,
\end{equation}
which is fully compatible with zero.

Obviously, in order to access physically 
sensible values of the coupling constant
[see Eq.~\eqref{epsEbound}],
\begin{equation}
2 \times 10^{-4} < \varepsilon_e \approx \varepsilon_\mu
< 1.4 \times 10^{-3} \,,
\end{equation}
one needs to increase the experimental precision.
In view of the inequality~\eqref{predestined}, 
muonic carbon appears to be well suited for an 
extraction of the $X$ boson contribution, based on 
spectroscopic data alone. The idea is to use the state dependence
of the $\xi$ parameter, in order to be able to 
write a non-singular system of the equations which can 
be solved for the nuclear radius and the coupling parameters
of the $X$ boson. 
Let us denote by $\nu_{1S\, 2P}$ 
and $\nu_{2S\, 2P}$ the remainder frequencies
obtained after subtracting all known relativistic 
and quantum electrodynamic (QED) contributions to the 
transition frequencies. Because the finite-size
effect and the $X$ boson Hamiltonian primarily shift 
$S$ states, one may write for the $nS$--$2P$ transition,
\begin{align}
\nu_{nS\, 2P} =& \;
\xi_{1S} \, \left< 1S \left| H^{(\mu)}_X \right| 1S \right> +
\left< 1S \left| H_{\rm NFS} \right| 1S \right> 
\nonumber\\[0.1133ex]
=& \; r^2 _{{}^{12}{\rm C}} \,
\frac23 \, \frac{(Z\alpha)^4 \, m_\mu^3}{n^3} 
\nonumber\\[0.1133ex]
& \; + \varepsilon_\mu \, \xi_{nS} \,
\frac{4 (A-Z) \, (Z\alpha)^3 \, \alpha \, \varepsilon_n}{m_X^2 \, n^3} \,.
\end{align}
We here ignore reduced-mass corrections.
The system of equations
\begin{subequations}
\label{syseq}
\begin{align}
\nu_{1S\, 2P} =& \; 
\xi_{1S} \, \left< 1S \left| H^{(\mu)}_X \right| 1S \right> +
\left< 1S \left| H_{\rm NFS} \right| 1S \right> \,,
\\[0.1133ex]
\nu_{2S\, 2P} =& \; 
\xi_{2S} \, \left< 2S \left| H^{(\mu)}_X \right| 2S \right> +
\left< 2S \left| H_{\rm NFS} \right| 2S \right> \,,
\end{align}
\end{subequations}
can be solved for $\varepsilon_\mu$ and $r_{{}^{12}{\rm C}}$,
because of $\xi_{1S} \neq \xi_{2S} \neq 1$.
The solution is
\begin{subequations}
\label{sol}
\begin{align}
\varepsilon_\mu =& \; 
\frac{(\nu_{1S\, 2P} - 8 \, \nu_{2S\,2P}) \,
m_X^2}{2 (A-Z) (Z\alpha m_\mu)^3 \,\alpha \varepsilon_n} \, f(\chi) \,,
\\[0.1133ex]
f(\chi) =& \; \frac{(1 + \chi)^4 \, (2 + \chi)^2}%
{\chi^2 \, [ \chi (4 + 3\chi) - 2] } \,,
\\[0.1133ex]
r^2 _{{}^{12}{\rm C}} =& \;
\frac{3 \, \nu_{1S\, 2P}}{2 (Z\alpha)^4 \, m_\mu^3} 
+ 
\frac{3 \, (\nu_{1S\, 2P} - 8 \, \nu_{2S\,2P})}%
{(Z\alpha)^4 \, m_\mu^3} \, g(\chi) \,,
\\[0.1133ex]
g(\chi) =& \; \frac{(1 + \chi)^4}{ 2 - \chi (4 + 3\chi) } \,,
\end{align}
\end{subequations}
where $\chi$ has been defined in Eq.~\eqref{defchi}.
Plugging in the parameters for ${}^{12}{\rm C}$
(see Ref.~\cite{An2004}),
one obtains for the sensitivity
\begin{equation}
\label{sense}
\delta \varepsilon_\mu \approx
31.234 \, \frac{ \delta (\nu_{1S\, 2P} - 8 \, \nu_{2S\,2P})}{\nu_{1S\, 2P}}
\approx 31.234 \, 
\frac{ \delta r^2_{{}^{12}{\rm C} }}{ r^2_{{}^{12}{\rm C}}} \,,
\end{equation}
where $\delta (\nu_{1S\, 2P} - 8 \, \nu_{2S\,2P})$
is the uncertainty with which $\nu_{1S\, 2P} - 8 \, \nu_{2S\,2P}$
could be determined experimentally.
Also, we should clarify that 
$\delta r^2_{{}^{12}{\rm C} }$ is the difference in the 
nuclear radii, determined from the two transitions separately,
assuming that one ignores the possible presence of the $X$ boson.
A comparison to recent determinations
of nuclear radii for simple atomic systems~\cite{PoEtAl2010,AnEtAl2013,PoEtAl2016}
reveals that an increase in the current 
experimental accuracy by about two orders of
magnitude will be sufficient to discern the 
$X$ boson from atomic spectroscopy.
For muonic magnesium, the sensitivity coefficient
in Eq.~\eqref{sense} changes according to
the replacement~$31.234 \to 58.515$.

Various generalizations of the system of equations~\eqref{syseq}
are possible. One obvious generalization would concern 
additional carbon isotopes such as ${}^{13}{\rm C}$,
for which the expansion coefficients are a little 
different. In this case, if one obtains a consistent result
for $\varepsilon_\mu$ from two different isotopes, 
this will serve as an independent confirmation of the 
result. Other generalizations would include combinations
of transitions in muonic systems ($\xi_{nS} \neq 1$)
with electronic bound systems, where $\xi_{nS}$ is 
nearly equal to unity, in view of the relation 
$\lambdabar_X \ll \lambdabar_H$ [see Eqs.~\eqref{LX} and~\eqref{LH}].
Also, generalizations to transitions involving the 
$3S$ state are straightforward [see Eq.~\eqref{xi3S}].

%
%
\section{Conclusions}
\label{sec5}

In this article, we have studied the $X$ boson~\cite{FeEtAl2016,FeEtAl2017}
from the point of view of atomic physics,
both in terms of possible connections to the proton 
and deuteron charge puzzles~\cite{PoEtAl2010,AnEtAl2013,PoEtAl2016}
(see Sec.~\ref{sec3}) as well as muonic 
bound systems~(see~Sec.~\ref{sec4}).
As outlined in Sec.~\ref{sec2}, the parameter 
range of the $X$ boson is energetically somewhat 
outside of the
range of atomic physics and therefore,
the particle is hard to detect
by pure atomic physics techniques.
This fact, in particular, explains why it has not 
been seen in atomic experiments, despite heroic
efforts of experimentalists to increase the precision 
of measurements in simple atomic systems
(see, e.g., Ref.~\cite{MaEtAl2013prl}).
In fact, the range of the $X$ boson interaction
somewhat overlaps with the atomic nucleus;
it can be characterized as an interaction 
present in some extended ``nuclear halo'' with a 
range of about $11.8 \, {\rm fm}$ [see Eq.~\eqref{LX}].

For interactions involving bound muons, one has to use
the Yukawa potential~\eqref{yukawaX} instead of
the Dirac-$\delta$ approximation~\eqref{HXmu}.
This, however, does not imply an electron-muon
nonuniversality; it simply means that the 
$X$ boson effect has to be evaluated differently for 
bound electrons as opposed to muons.
The same phenomenon is observed (for 
electronic systems) with vacuum polarization,
where a good approximation is formed by a Dirac-$\delta$ 
potential for ordinary hydrogen, 
but one has to carry out a detailed
integration for muonic systems 
(see Ref.~\cite{Je2011aop1}), because 
the length scale of the bound muonic system 
is commensurate with the electron Compton 
wavelength, which in turn defines the 
extent of the vacuum-polarization mediated modification
of the Coulomb interaction.

This latter observation leads to a possible 
pathway toward the observation of the $X$ boson 
in atomic systems, as described in Sec.~\ref{sec4}.
A model calculation involving muonic carbon
illustrates that a nontrivial dependence of 
the $X$ boson effect on the principal quantum 
number is introduced for $S$ states in 
muonic systems, which leads to a 
separation of the effect from the nuclear-size contribution,
rendering the $X$ boson effect observable 
[see Eq.~\eqref{sol}].

\acknowledgements

The authors acknowledge insightful discussions
with Prof.~A.~Krasznahorkay.
Support by the National Science Foundation
(grants No.~PHY--1403973 and No.~PHY--1710856)
also is gratefully acknowledged.
A Missouri Research Board grant also assisted 
the completion of this research,
and the work was supported by a J\'{a}nos Bolyai 
Research Scholarship of the
Hungarian Academy of Sciences.

\appendix

%
%
\section*{Appendix: Couplings in the Neutrino Sector}

This brief appendix is devoted to the 
discussion of the $X$ boson model in a particle 
physics context, with a particular emphasis 
on the neutrino sector. We recall that in 
Eq.~\eqref{LL}, the couplings to the fermion fields are left as free parameters 
in the $X$ boson coupling Lagrangian.
In Sec.~\ref{sec1}, we have discussed constraints on these parameters 
for electrons, protons and neutrons, the latter 
being determined according to their 
quark content~\cite{FeEtAl2016,FeEtAl2017}. 

Important constraints on the coupling parameters for neutrinos
have been discussed in Sec.~VI.C
of Ref.~\cite{FeEtAl2017}.
Namely, according to Sec.~VI.C.1 of Ref.~\cite{FeEtAl2017},
some of the most stringent constraints 
come from the TEXONO experiment,
where electron (anti-)neutrinos scatter off electrons.
Because of a relatively small length of the interaction 
region (of about 28 meters), the 
electrons (of energy $1$--$2$\,MeV) 
remain in pure electronic flavor eigenstates.

Depending on the sign of the coupling parameters
of electrons and neutrinos, the interference 
of the $X$ boson term can lead to constructive
or destructive interference with the Standard Model 
prediction. According to Sec.~VI.C.1 of Ref.~\cite{FeEtAl2017},
for the electron coupling parameter range given in 
Eq.~\eqref{epsEbound}, 
one finds bounds for $|\varepsilon_\nu|$ in the range
from $10^{-6}$ to $10^{-4}$
for constructive and destructive interference alike.
Here, $\varepsilon_\nu$ is the electron (anti-)neutrino
coupling parameter.

Neutrino-nucleus scattering has not yet been observed,
but it is the target of a number of 
upcoming experiments that use reactors as sources. 
According to Sec.~VI.C.2 of Ref.~\cite{FeEtAl2017},
from SuperCDMS, CDMSlite, and LUX, one obtains bounds 
for $|\varepsilon_\nu|$ in the range
from $10^{-5}$ to $10^{-4}$ for the electron neutrino 
coupling parameter $\varepsilon_\nu$,
assuming that $|\varepsilon_n| = 1/100$.
These constraints are not in disagreement with any 
other experimental observations.

Interesting connections to the
neutrino sector have also been 
pointed out in Ref.~\cite{Eh2017model}, where 
a dark matter particle $D$ with mass $8.4$\,MeV is 
being proposed, 
which would give rise to the
reaction $D + D \to X$, where the $X$ particle has a 
predicted mass of 
$16.8$\,MeV, just twice the $D$ mass, 
almost perfectly matching the proposed
$X$ boson mass~\cite{FeEtAl2016,FeEtAl2017}.
The $D$ particle is 
required for the interpretation of the
Mont Blanc neutrino burst~\cite{DaEtAl1987},
as proposed in Ref.~\cite{Eh2017model}.

In Ref.~\cite{SeSh2017} (see also Ref.~\cite{KaSeSh2011}),
the authors identify the $X$ boson as the massive
vector boson of a new $U(1)$ gauge group, which, by virtue of 
the interaction Lagrangian [see Eq.~(2) of Ref.~\cite{SeSh2017}],
is called a baryon minus lepton ($B-L$) symmetry.
In addition to explaining the ATOMKI 
anomaly~\cite{KrEtAl2016,*KrEtAl2017woc2,Ca2016news,Ta2016}, 
the $U(1)_{B-L}$ also provides a possible explanation for the lightness
of the neutrinos, by proposing a 
radiative seesaw model in which 
neutrinos acquire their tiny masses only by a one-loop
diagram whose value is proportional to 
the vacuum expectation value $v_s$ of a
scalar field $S$ which takes the role of 
an added Higgs-like particle 
[see Eq.~(5) of Ref.~\cite{SeSh2017}].
Likewise, the mass of the $X$ boson is proportional 
to $v_s$ [see Eq.~(12b) of Ref.~\cite{SeSh2017}].
In the context of the  $U(1)_{B-L}$ models,
the authors of Ref.~\cite{FaHe2016} point out that it could be quite natural
to assume a protophobic interaction
($\varepsilon_p \simeq \varepsilon_e \ll 1$), 
but then, it would be more natural to 
assume that the couplings to neutrinos 
are not as suppressed as indicated in Sec.~VI.C of Ref.~\cite{FeEtAl2017},
but rather, that $\varepsilon_n \simeq -\varepsilon_\nu$.
Finally, according to Ref.~\cite{KaEtAl2016atomki},
the new X boson could also
help in resolving a $2$--$3$\,$\sigma$ discrepancy between 
theory~\cite{DoIv2007} and experiment~\cite{AbEtAl2007}
for the rare decay $\pi^0 \to e^+\,e^-$.

\color{black}

\end{document}